\documentclass[twocolumn,aps,prb,showpacs]{revtex4}
\usepackage{amsmath} 
\usepackage{amssymb} 
\usepackage{arydshln}
\usepackage{graphics}
\usepackage{epsfig}
\usepackage{epstopdf}
\usepackage{color}
 \usepackage{rotating}
 \usepackage{bm}
 
\begin{document}
\title{Spin liquid phase in a spatially anisotropic frustrated antiferromagnet}  
\author{Jaime Merino}
\affiliation{Departamento de F\'isica Te\'orica de la Materia Condensada, Condensed Matter Physics Center (IFIMAC) and Instituto Nicol\'as Cabrera, Universidad Aut\'onoma de Madrid, Madrid 28049, Spain} 
\author{Michael Holt}
\author{Ben J. Powell}
\affiliation{School of Mathematics and Physics, University of Queensland,
  Brisbane, 4072 Queensland, Australia} 
\date{\today}
                   
\begin{abstract}
We explore the effect of the third nearest-neighbors on the magnetic properties of the 
Heisenberg model on an anisotropic triangular lattice. We obtain the phase diagram of the model 
using Schwinger-boson mean-field theory. Competition between N\'eel, spiral and collinear 
magnetically ordered phases is found as we vary the on the ratios of the nearest, $J_1$,
next-nearest, $J_2$, and third-nearest, $J_3$, neighbor  exchange 
couplings. A spin liquid phase is stabilized between the spiral 
and collinear ordered states when
$J_2/J_1 \gtrsim 1.8$ for rather small $J_3/J_1 \lesssim 0.1$. 
The lowest energy two-spinon dispersions 
relevant to neutron scattering experiments are analyzed and compared to 
semiclassical magnon dispersions finding significant differences in the spiral and 
collinear phases between the two approaches. The results are discussed in the context 
of the anisotropic triangular materials: Cs$_2$CuCl$_4$ and Cs$_2$CuBr$_4$ and 
layered organic materials, $\kappa$-(BEDT-TTF)$_2X$ and $Y$[Pd(dmit)$_2$]$_2$. 
\end{abstract}

\maketitle 
 
\section{Introduction}
\label{sec:intro}


Quantum spin liquids (QSL) are exotic states of matter with no broken symmetries even 
at zero temperature \cite{balents2010}. Fractional excitations such as deconfined spin $S=1/2$ 
spinons are expected to occur as well as emergent gauge fields. These exotic phenomena   
are typically explored in low-dimensional $S=1/2$ systems. However, 
understanding the precise conditions for the realization of a QSL is a major 
challenge in theoretical condensed matter physics. 
For instance, in the one-dimensional $S=1/2$ Heisenberg model, low
energy  magnetic excitations are not the conventional $S=1$ magnons expected in an ordered magnet\cite{ashcroftbook}
but $S=1/2$ spinons which propragate as domain walls along the chain.\cite{Tsvelik} While this is a well understood 
example of fractionalization, the existence of such fractional excitations in 
a two-dimensional spin system remains unsettled.

As well as the fundamental theoretical interest further impetus to investigate QSLs 
has arisen from recent experimental observations 
identifying several materials in which such unconventional behavior may be realized.  
The $\kappa$-(BEDT-TTF)$_2X$ and $Y$[Pd(dmit)$_2$]$_2$
families of organic charge transfer salts include spin liquid materials such as
$\kappa$-(BEDT-TTF)$_2$Cu$_2$(CN)$_3$ and Me$_3$EtSb[Pd(dmit)$_2$]$_2$ 
where Et = C$_2$H$_5$ and Me= CH$_3$ in contrast to other antiferromagnetically 
ordered Mott insulators such as the $X =$Cu[N(CN)$_2$]Cl salts. There have also been  predictions of a spin liquid in Mo$_3$S$_7$(dmit)$_3$, where the molecules themselves provide a triangular motif.\cite{Janani} There are also a number of possible spin liquids in inorganic materials. Cs$_2$CuCl$_4$, does not display spiral magnetic order 
down to $T=0.62$ K and  Cs$_2$CuBr$_4$ is also a candidate system for spin liquid behavior. 
Both the organic and inorganic materials discussed above have been primarily modeled in terms 
of the Heisenberg model on an anisotropic triangular lattice with exchange constants $J_1$ and $J_2$. The organic materials 
$\kappa$-(BEDT-TTF)$_2$Cu$_2$(CN)$_3$ and Me$_3$EtSb[Pd(dmit)$_2$]$_2$ are in the regime \cite{scriven2012}, $J_2/J_1 \approx 0.7 $, 
whereas Cs$_2$CuCl$_4$   ($J_2/J_1 \approx 3$)  and  Cs$_2$CuBr$_4$  ($J_2/J_1 \approx 2 $) are 
closer to the weakly coupled chain limit \cite{john2007}. Other materials which may 
display spin liquid behavior are Ba$_3$CoSb$_2$O$_9$ (Ref. \onlinecite{shirata2012}) and Ba$_3$CuSb$_2$O$_9$ (Ref. \onlinecite{zhou2011})
which have isotropic triangular lattices, \cite{susuki2013} $J_2/J_1=1$.  

There are several experimental observations which suggest the existence of spin liquid behavior 
in these materials. Susceptibility and NMR measurements in $\kappa$-(BEDT-TTF)$_2$Cu$_2$(CN)$_3$ and Ba$_3$CuSb$_2$O$_9$ 
find no magnetic order down to very low temperatures,\cite{RPP} much lower than $J_1$.  The 
specific heat probing magnetic excitations reveals a linear 
temperature dependence\cite{RPP} in such Mott insulators suggesting  the existence of 
a Fermi surface consisting of fractional excitations (spinons)\cite{sachdev1992,motrunich2005}.  
In $\kappa$-(BEDT-TTF)$_2$Cu$_2$(CN)$_3$, a power-law $T$-dependence, $1/T_1 \propto T^{3/2}$ below 1 K \cite{shimizu2006} is
observed. The absence of magnetic order together with the power-law $T$-dependence suggest the vanishing of the  gap to triplet excitations.\cite{Normand,RPP}
NMR experiments on Cs$_2$CuCl$_4$ show \cite{marston2012, balents2007} a linear dependence of 
the relaxation rate with temperature $1/T_1 \propto T$ in the short range ordered region $T>0.62$ K. In the same temperature regime neutron scattering experiments observe a continuum of excitations constant with the presence of deconfined spinons.\cite{coldea2001}


The above unconventional behavior is difficult to understand theoretically.
For instance, there is overwhelming numerical evidence that the Heisenberg model on an isotropic 
triangular lattice has the 120$^0$ N\'eel ordered state \cite{elstner1993,bernu1994} as
the ground state in contrast to Anderson's original prediction for a 
spin liquid\cite{anderson1973}. This seems consistent with the antiferromagnetic (AF) order observed in the
nearly isotropic organic materials\cite{scriven2012}: Me$_4$Sb[Pd(dmit)$_2$]$_2$ and 
Me$_2$Et$_2$As[Pd(dmit)$_2$]$_2$. However it is inconsistent with observations in 
isotropic triangular lattice materials: Ba$_3$CoSb$_2$O$_9$ \cite{shirata2012} and Ba$_3$CuSb$_2$O$_9$. 
Hence, other interaction terms not present in the nearest-neighbor Heisenberg 
model should be included to explain discrepancies with the observations. \cite{scriven2012}

One possible route to 
spin liquid behavior is the presence of further neighbor AF exchange couplings not considered
in the nearest-neighbor models. These can be generated
through the, second order, superexchange mechanism, {\it i.e.}, $J_3 \propto t_3^2/U$, where $t_3$ is the hopping integral between third neighbouring sites, between 
the third-nearest neighbor sites. 
Alternatively fourth order process can give rise a $J_3 \propto t_1^2t_2^2/U^3$, where $t_1$ and $t_2$ are the nearest and next nearest neighbour hopping integrals. These fourth order process also give rise to a ring exchange term, $J_3 ({\bf S}_i\cdot {\bf S}_j)({\bf S}_k\cdot {\bf S}_l)$, where ${\bf S}_i$ is the Heisenberg spin operator on the $i^{th}$ site.

Second and third nearest neighbor   AF exchange coupling frustrates magnetically ordered phases 
and can lead to spin liquid behavior. For instance, Wang and Vishwanath \cite{wang2006} have found 
spin disordered flux phases in the large quantum fluctuation regime when $S<1/2$. 
Ring exchange can also lead to spin liquid behavior on the isotropic triangular \cite{liming2000,motrunich2005}
and anisotropic triangular lattices. \cite{michael2013}
 On the isotropic triangular 
lattice the two contributions generated by four order processes lead to spin liquid 
behavior (for $J_3/J_1>0.1$) which is characterized by gapless magnetic excitations and a spinon Fermi 
surface.\cite{motrunich2005} It is then interesting to understand the effect of each 
contribution separately. Alternatively, other mechanisms may also stabalize spin liquids. For example, it has been argued that the Dzyakoshinskii-Moriya interaction may also  
produce a spin liquid behavior in Kagom\'e lattices \cite{cepas2010} and anisotropic 
triangular lattices.\cite{john2007}

The main aim of the present work is to analyze the effect of next-nearest neighbor interaction, $J_3$,  on 
the magnetic properties of the Heisenberg model on anisotropic triangular 
lattices. Since these interactions can be generated by fourth order process that also lead to ring exchange as discussed above, 
our work contributes to the general understanding of ring exchange effects on frustrated 
antiferromagnets \cite{michael2013}. We use Schwinger boson mean-field theory (SB-MF)\cite{auerbachbook} 
expressed in terms of antiferromagnetic and ferromagnetic bonds which are treated as 
variational parameters\cite{gazza1993,mezio2012,flint2009}. The Schwinger boson approach is particularly
useful since it can describe ordered and disordered phases on equal footing; the magnetically ordered
phases resulting from the condensation of the bosons at particular order wavevectors of the system.
We find that when the anisotropy $J_2/J_1 \gtrsim 1.8$ of the system is amenable to 
spin liquid behavior under the effect of a weak next-nearest neighbor interaction, $J_3/J_1 \lesssim 0.1$. 
Since these results are obtained from Schwinger boson mean-field theory which favors broken 
symmetry magnetic phases, our results suggest that the    
spin liquid phase found here is robust against fluctuations. 
This spin liquid discussed below is most relevant to the spin liquid candidate materials typically 
modelled through anisotropic triangular lattices with $J_2/J_1>1$ such as 
Cs$_2$CuCl$_4$ and Cs$_2$CuBr$_4$ for which the third nearest-neighbor interactions 
are typically neglected.


The present paper is organized as follows: in section \ref{sec:model}, the $J_1-J_2-J_3$ 
Heisenberg model studied is introduced. In section \ref{sec:sbmf}, the Schwinger boson 
formulation is briefly revised and main issues described. In section \ref{sec:phased}, the ground state energies,
magnetization and phase diagram obtained with SB-MF are obtained and discussed. Elementary magnetic
excitations of the system are discussed in Section \ref{sec:spinons}.  We finally 
end up with conclusions and the relevance to anisotropic triangular lattice materials in section \ref{sec:conclusions}.

\section{Heisenberg model on an anisotropic triangular lattice with third-nearest neighbor interactions}
\label{sec:model}

\begin{figure}
\epsfig{file=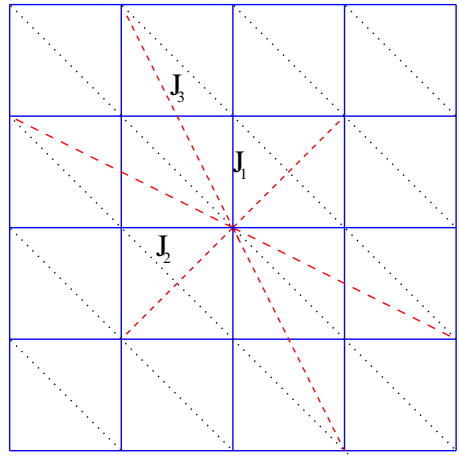,width=6.cm,angle=0,clip=}
\caption{(Color online) Picture of the $J_1-J_2-J_3$ Heisenberg model (\ref{eq:model}) considered. The 
lattice model studied has the same topology as the original anisotropic triangular lattice model in 
which each lattice site is connected to its nearest, next-nearest and third nearest neighbor
sites through the antiferromagnetic couplings: $J_1$, $J_2$ and $J_3$, respectively.}
\label{fig:model}
\end{figure} 

 We are interested in understanding the magnetic properties of the Heisenberg model on the anisotropic 
 triangular lattice including exchanges up to third nearest-neighbor spins: 
\begin{equation}
H=J_1 \sum_{\langle ij \rangle} {\bf S}_i \cdot {\bf S}_j + J_2 \sum_{\langle\langle ij \rangle\rangle}{\bf S}_i \cdot {\bf S}_j
+J_3\sum_{\langle\langle\langle ij \rangle\rangle\rangle} {\bf S}_i \cdot {\bf S}_j.
\label{eq:model}
\end{equation}
We take from now on $J_1=1$ unless otherwise stated.  Sum $\langle ij \rangle$ runs over nearest-neighbors, 
$\langle\langle ij \rangle\rangle$ runs over next nearest-neighbors and $\langle\langle\langle ij \rangle\rangle\rangle$ 
over third nearest-neighbor pairs of sites. 
The anisotropic triangular lattice model with no third-nearest neighbors, $J_3=0$, has been studied extensively.\cite{RPP}
Related models including ring exchange contributions also have been recently analyzed.\cite{michael2013}
For the particular case of the isotropic triangular case, $J_2=1$ and $J_3=0$, Sachdev \cite{sachdev1992} finds a spin liquid
phase which becomes the long range 120$^0$ magnetically ordered state when the
quantum fluctuations are reduced to $S=1/2$ within a Sp$(N)$ formulation of the Heisenberg model where
$N$ is the number of spin species. 
The general $J_2\neq 1$ situation has been explored using exact diagonalization and DMRG techniques  \cite{bursill2006} 
linear spin-wave theory (LSWT) \cite{trumper1999,merino1999}, modified spin-wave theory \cite{hauke2013},
series expansions\cite{zheng1999}, 
mean-field Schwinger boson theory \cite{manuel1999} and large-$N$ approaches\cite{chunghou2001}. 
In the region where a transition from N\'eel antiferromagnetism to spiral order occurs ($J_2=0.5$ within LSWT) 
a spin liquid has been speculated to exist. 
The isotropic triangular lattice model, $J_2=1$, under the effect of  $J_3$
has been studied \cite{gazza1993} using Schwinger boson mean-field theory and recently 
revisited \cite{wang2006}.  
Spin liquid phases have recently been found in the Hubbard model on the anisotropic 
triangular lattice.  \cite{tocchio2013}

\begin{figure}
\epsfig{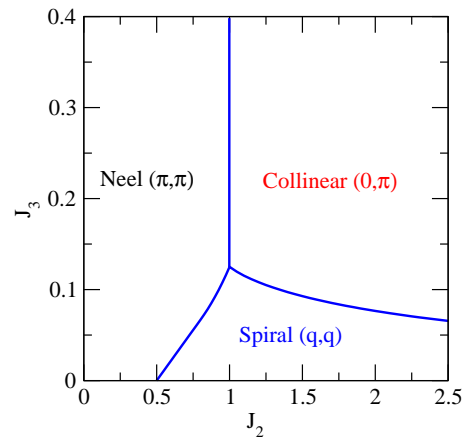}
\caption{(Color online) Classical phase diagram for the  
$J_1-J_2-J_3$ Heisenberg model (\ref{eq:model}) for the helical ground states. We have taken $J_1=1$.
}
\label{fig:phasedclass}
\end{figure}

{\it Classical limit:} the classical ground state energy 
 of model (\ref{eq:model}) is evaluated considering planar helices only. The
spin at each site is given by: ${\bf S_i}=S\cos({\bf Q}\cdot {\bf R}_i) {\bf e_1} + S\sin({\bf Q}\cdot {\bf R}_i)
{\bf e_2}$, ${\bf e}_1$ and ${\bf e}_2$ being an orthonormal basis and ${\bf Q}=(Q_x,Q_y)$ the ordering
wavevector. The classical phase diagram is obtained by comparing the energies of the spiral 
(${\bf Q}=(Q,Q)$), collinear (${\bf Q}=(0,\pi)$) and N\'eel (${\bf Q}=(\pi,\pi)$) orders. The 
wave vector of the spiral phase is given by:
\begin{equation}
Q=\arccos \left( { -J_2+\sqrt{J_2^2+12J_3(3J_3-1)}  \over 12 J_3 } \right).
\label{eq:Q}
\end{equation} 
The phase diagram resulting from these three phases is shown in Fig. \ref{fig:phasedclass}. For 
$J_3 \rightarrow 0$ the transition between N\'eel and spiral order with $Q=\arccos(-1/2J_2)$ occurs at $J_2=0.5$
as expected for the anisotropic triangular lattice. In the isotropic limit, $J_2=1$,
the transition from the spiral to collinear or N\'eel orders occurs at $J_3=1/8$. This is in agreement  
with the spin wave analysis of model (\ref{eq:model}) on the isotropìc triangular lattice: $J_1=1$, $J_2=1$ and $J_3=0$  
\cite{joliceour1990}.     

\section{Schwinger boson mean-field theory}
\label{sec:sbmf}

The quantum magnetism of bipartite (unfrustrated) lattices
can be explored using the Schwinger bosonic representation of $SU(N)$
Heisenberg models\cite{auerbachbook}. Extensions to frustrated lattices 
can be done by \cite{read1991} using the Sp$(N)$ representation. 
Here, we use the SU(2) mean-field theory introduced by 
Cecatto {\it et. al.} \cite{ceccatto1993} which keeps ferromagnetic 
and antiferromagnetic components in the mean-field approach.  
Such mean-field decoupling was found to correspond to the large-$N$
limit of a "symplectic-N" representation of the spins \cite{flint2009} which 
appropriately takes into account time-reversal properties of the spins in frustrated magnets.
The Schwinger boson approach can describe both magnetically ordered
and disordered states complementing other semiclassical spin-wave 
theories.  We now summarize the main steps in the Schwinger mean-field approach 
\cite{auerbachbook} to the Heisenberg model (\ref{eq:model}) following previous works 
\cite{ceccatto1993,gazza1993,lefmann1994,flint2009,mezio2011,mezio2012}. 
 
Schwinger bosons are used to express the Heisenberg interaction terms in the model (\ref{eq:model}).
Each bond between two different sites is expressed through the operator identity:
 \begin{equation}
 {\bf S}_i\cdot {\bf S}_j=:\hat{B}_{ij}^\dagger \hat{B}_{ij}:-\hat{A}_{ij}^\dagger \hat{A}_{ij},
 \end{equation}
 where $::$ is normal ordering, and the operators $\hat{A}_{ij}$ and $\hat{B}_{ij}$ are defined in terms 
 of the Schwinger bosons as: 
 \begin{eqnarray}
 \hat{A}_{ij} &=& {1 \over 2} ( a_{i\uparrow} a_{j\downarrow}-a_{i\downarrow} a_{j\uparrow} )
 \nonumber \\
 \hat{B}_{ij} &=& {1 \over 2} ( a^\dagger_{i\uparrow} a_{j\uparrow}+a^\dagger_{i\downarrow} a_{j\downarrow} ),
 \end{eqnarray}
 where $a^\dagger_{i\uparrow}$ and  $a^\dagger_{i\downarrow}$ create a "spin up"  and 
 "spin down" Schwinger boson on site $i$. The two operators, $\hat{A}_{ij}$ and $\hat{B}_{ij}$ 
 describe antiferromagnetic and ferromagnetic bonds between $i$ and $j$ sites, respectively. 
 
 The magnitude of the spin is fixed by restricting the number of bosons per site:
 \begin{equation}
 \sum _\sigma a_{i\sigma}^\dagger a_{i\sigma}=2S,
 \label{eq:constraint}
 \end{equation}
which is the constraint equation imposed over the Schwinger bosons avoiding 
having an arbitrary number of bosons at each site.

After a mean-field decoupling of the quartic terms describing the bonds, the Heisenberg model 
(\ref{eq:model}) can be expressed as a quadratic hamiltonian:
 \begin{eqnarray}
 H&=&J_1\sum_{\langle ij \rangle } (B_{ij}^*\hat{B}_{ij} - A^*_{ij}\hat{A}_{ij} +H.c.)
 \nonumber \\
 &+&J_2\sum_{\langle \langle ij  \rangle \rangle } (B_{ij}^*\hat{B}_{ij} - A^*_{ij}\hat{A}_{ij} +H.c.)
 \nonumber \\
 &+&J_3\sum_{\langle \langle \langle ij \rangle \rangle \rangle } (B_{ij}^*\hat{B}_{ij} - A^*_{ij}\hat{A}_{ij} +H.c.)
 \nonumber \\
&+&J_1\sum_{\langle ij \rangle } (-B_{ij}^*{B}_{ij} +A^*_{ij}A_{ij} )
\nonumber \\
&+&J_2\sum_{\langle \langle ij  \rangle \rangle } (-B_{ij}^*B_{ij} +A^*_{ij}A_{ij} )
\nonumber \\
&+&J_3\sum_{\langle \langle \langle ij \rangle \rangle \rangle } (-B_{ij}^*B_{ij} +A^*_{ij}A_{ij} )
\nonumber \\
&+&\lambda \sum_i(\sum_\sigma \langle a^+_{i\sigma} a_{i\sigma} \rangle -2S).
 \end{eqnarray}
 The variational energy of the system is minimized with respect to $A_{ij}$
 and $B_{ij}$ and the Lagrange multiplier $\lambda$ fixes the constraint (\ref{eq:constraint})
 at each site on average. The resulting set of self-consistent equations obtained are:
 \begin{eqnarray}
 \langle \hat{A}_{ij}\rangle=A_{ij}, \nonumber \\ 
 \langle \hat{B}_{ij}\rangle=B_{ij},  \nonumber \\
\sum _\sigma \langle a_{i\sigma}^\dagger a_{i\sigma} \rangle =2S,
 \end{eqnarray}
and the variational bond energy reads:
 \begin{equation}
 \langle {\bf S}_i\cdot {\bf S}_j\rangle =|B_{ij}|^2-|A_{ij}|^2.
 \end{equation}
 After Fourier transformation, the mean-field hamiltonian reads:
 \begin{eqnarray}
 H^{MF}&=&\sum_{{\bf k},\sigma} (B({\bf k}) + \lambda) a^\dagger_{{\bf k},\sigma} a_{{\bf k},\sigma}\\\notag&& -i \sum_{\bf k} A({\bf k}) (a_{{\bf k} \uparrow} a_{-{\bf k}\downarrow }
 +a^\dagger_{{\bf k} \uparrow} a^\dagger_{-{\bf k}\downarrow }) -2\lambda N_s S,
 \end{eqnarray}
 with $N_s$ the number of sites in the lattice. The coefficients $A({\bf k})$ and $B({\bf k})$ are given by:
 \begin{eqnarray}
  A({\bf k})&=&{1 \over 2} \sum_{{\bm{\delta}}_i} J_i  \sin({\bf k} \cdot {\bm{\delta}}_i) A_{\bm{\delta}_i},
  \nonumber \\
  B({\bf k})&=&{1 \over 2}  \sum_{{\bm{\delta}}_i}J_i  \cos({\bf k} \cdot {\bm{\delta}}_i) B_{\bm{\delta}_i}
 \label{eq:MF}
  \end{eqnarray}
 where the sums are performed over the $\bm{\delta}_i$ vectors connecting pairs of sites coupled  
 by $J_i$; {\it i. e.} $\delta_1$ refers to the vector connecting nearest neighbor, $\delta_2$ next-nearest neighbor
and $\delta_3$ third nearest-neighbor sites. 
 The variational parameters satisfy:  $A_{-{\bm{\delta}}_i}=-A_{{\bm{\delta}}_i}$ and 
 $B_{-{\bm{\delta}}_i}=B_{{\bm{\delta}}_i}$, when evaluating the sums over $\bm{\delta}_i$. 

A Bogoliubov transformation is performed to diagonalize the hamiltonian. This leads to the following mean-field 
hamiltonian: 
\begin{equation}
H^{MF}=\sum_{{\bf k}, \sigma} \omega({\bf k}) (\alpha^\dagger_{{\bf k}, \sigma} \alpha_{{\bf k}, \sigma} + {1 \over 2} )
-N_s\lambda (1+2S).
\end{equation}
where the Bogoliubov quasiparticle operator is expressed as: $\alpha^\dagger_{{\bf k},\sigma}=\cosh(\theta_{\bf k}) 
a^\dagger_{{\bf k},\uparrow} -\sinh(\theta_{\bf k}) a_{-{\bf k}\downarrow}$, in terms of the original bosons 
with: $\tanh(\theta_{\bf k})=-{A({\bf k}) \over B({\bf k})+\lambda}$. These Bogoliubov quasiparticles have 
the following dispersion:
\begin{equation}
\omega({\bf k}) = \sqrt{(B({\bf k})+\lambda)^2-A({\bf k})^2}.
\label{eq:dispersion}
\end{equation}
From the minimization of the total energy, $E_0= \langle H^{MF} \rangle $, a set of self-consistent equations: 
\begin{eqnarray}
{1 \over 2N_s} \sum_{\bf k} { A({\bf k} ) \over \omega({\bf k}) } \sin({\bf k} \cdot {\delta_i})= A_{{\bf \delta}_i}
\nonumber \\
{1 \over 2N_s} \sum_{\bf k} { B({\bf k} )+ \lambda \over \omega({\bf k}) } \cos({\bf k}  \cdot {\delta_i})= B_{{\bf \delta}_i}
\nonumber \\
{1 \over 2N_s} \sum_{\bf k} { B({\bf k} )+ \lambda \over \omega({\bf k}) } = {1 \over 2}+S,
\label{eq:mf}
\end{eqnarray}
are obtained at temperature $T=0$, which are numerically solved. 

%
%
In a finite lattice with $N_s$ sites magnetic ordering with a particular  order  is signalled by a 
minimum gap  in the spinon dispersion (located at $\pm {\bf Q}/2$) which scales as: 
$\omega_{\pm {\bf Q}/2} \sim 1/N_s$, scaling to zero with the system size.  In the thermodynamic limit, 
these modes go to zero and Bose condensation occurs at these wave vectors which
signals a magnetically ordered state with ordering vector, ${\bf Q}$. In infinite lattices, the sums 
in Eq. (\ref{eq:mf}) are converted into integrals separating the macroscopic contribution
of the condensed boson fraction at  $\pm {\bf Q}/2$,  which is treated as a self-consistent parameter, $m({\bf Q})$. 
The self-consistent equations (\ref{eq:mf}) are solved under the extra condition:  
$\omega_{{\bf Q}/2}=0$, which fixes $\lambda=A({\bf Q}/2)-B({\bf Q}/2)$ at each iteration.  

In large but finite systems, the magnetization can be obtained from: \cite{mezio2011}
\begin{equation}
m({\bf Q})={1 \over N_s} {  B({\bf Q}/2)+\lambda \over \omega({\bf Q}/2) }.
\end{equation}

We have checked that the magnetization, $m({\bf Q})$, and
total energy, $E_0$, converge to the thermodynamic limit results as the number of sites, $N_s$, is increased.
One can show that the classical ground state energy is recovered by SB-MF\cite{gazza1993} in the $S\rightarrow \infty$
limit as it should (see Appendix for details).

\section{Ground state properties}
\label{sec:phased}

We now analyze the ground state properties of the Heisenberg model (\ref{eq:model}). 
We first discuss the phase diagram of the anisotropic triangular lattice and then 
the effect of the third nearest-neighbor interactions, $J_3$, on the phase diagram. 

\subsection{Anisotropic triangular lattice model ($J_3=0$)}

\begin{figure*}
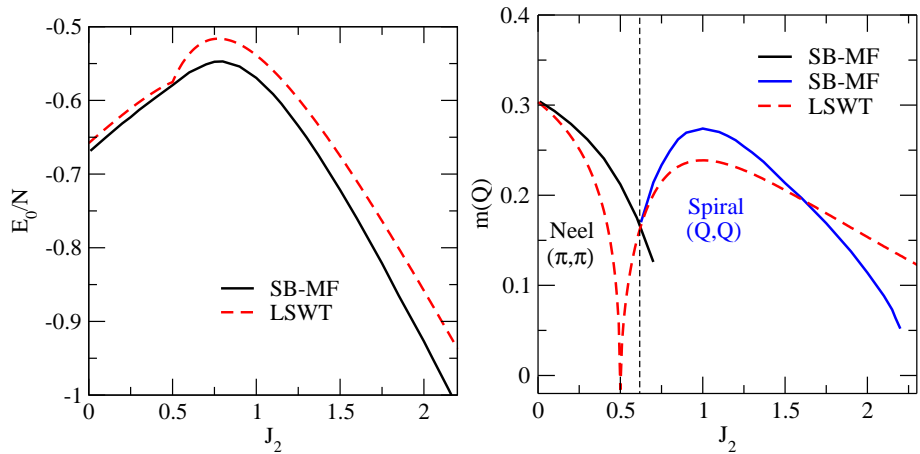

\epsfig{file=fig3a.eps,width=6cm,angle=0,clip=}
\epsfig{file=fig3b.eps,width=6cm,angle=0,clip=}
\caption{(Color online) Ground state properties of the anisotropic triangular lattice. 
The ground state energy (left) and magnetization (right) of model (\ref{eq:model}) with 
$J_3=0$ are shown. Dependence of ground state energy, $E_0$ and magnetization, $m({\bf Q})$, 
with $J_2$ from Schwinger boson mean-field theory. Schwinger boson mean-field theory does not show a 
disordered spin liquid phase between the N\'eel and spiral phase 
in contrast to spin-wave theory (dashed lines) at $J_2=0.5$. A spin liquid phase occurs in SB-MF 
for $J_2>2.2$, a much smaller value than that obtained from series expansions or LSWT. The dotted vertical line 
marks the onset of the continuous direct transition from the N\'eel 
to the spiral pase in SB-MF for $J_2 \approx 0.63$.\cite{manuel1999}}
\label{fig:moment}
\end{figure*}

It is illustrative to analyze first the ground state properties of the anisotropic triangular lattice ($J_3=0$)
with the SB-MF approach. In Fig. \ref{fig:moment} we plot the $J_2$ dependence of magnetization and total energy.  
The magnetic wave vector, ${\bf Q}$, changes continuously from $(\pi,\pi)$ to $(Q,Q)$ \cite{manuel1999} 
at $J_2 \approx 0.62$, which is larger than the classical transition point: $J_2=0.5$ with
no disordered phase found between N\'eel and spiral phases. Although the shift to higher $J_2$ critical 
values than the classical ones is consistent with series expansion \cite{zheng1999} results, the SB-MF  
fails to describe the disordered region around $0.7<J_2< 0.9$ or 
the disordered phase at $J_2=0.5$ predicted by linear spin-wave theory (LSWT)  
\cite{trumper1999,merino1999}. On the other hand, increasing $J_2 \approx 2.2$ a transition to a 
disordered state occurs consistent with the expected spin liquid phase in decoupled $S=1/2$ spin chains ($J_1 =0$). 
Note that this critical SB-MF value is much smaller than $J_2 \approx 3.8$ from LSWT\cite{trumper1999,merino1999}
or series expansions, $J_2 \approx 4.5$.\cite{zheng1999}    

\subsection{Effect of third nearest-neighbor interactions ($J_3 \neq 0$)}

\begin{figure*}
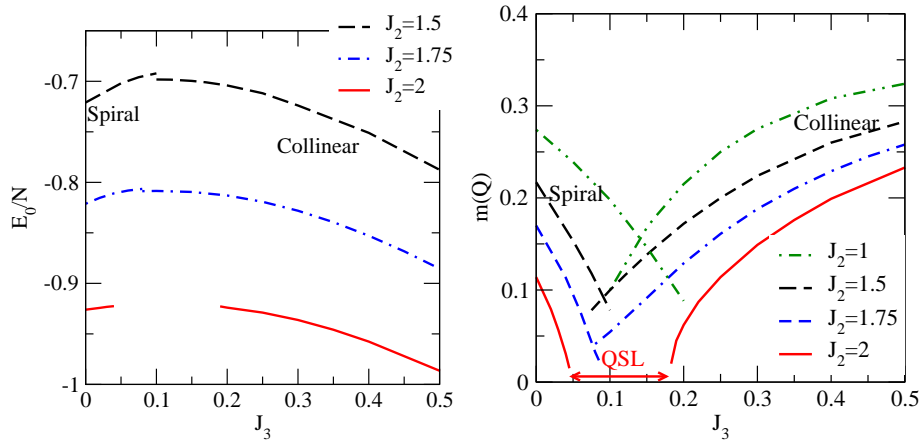
 
\epsfig{file=fig4a.eps,width=6cm,angle=0,clip=}
\epsfig{file=fig4b.eps,width=6cm,angle=0,clip=}
\caption{(Color online) Spin liquid phase in the $J_1-J_2-J_3$ Heisenberg model on the 
anisotropic triangular lattice. The 
dependence of magnetization and energy on the third nearest-neighbors
interaction, $J_3$, from Schwinger boson mean-field theory in infinite lattices. The energy curves are broken
in the region where no magnetically ordered solution is found. A spin liquid (QSL) phase occurs between
the spiral-$(Q,Q)$ and collinear-$(0,\pi)$ phase for $J_2 \gtrsim 1.8$ for a small $J_3 <0.1$. }
\label{fig:momentJ1J2J3}
\end{figure*}

We now analyze the effect of the third nearest-neighbor interaction. Results for 
the total energy per site and magnetization dependence on $J_3$ 
are shown in Fig. \ref{fig:momentJ1J2J3} for different $J_2$.  
 
The isotropic triangular lattice case has been previously studied\cite{gazza1993} with Schwinger
bosons and recently revisited \cite{wang2006}.
A first-order transition from 120$^0$-N\'eel ordering (${\bf Q}=(2\pi/3, 2\pi/3)$)  
to collinear order with ${\bf Q}=(0,\pi)$ occurs at about $J_3 \approx 0.16$.
These values should be compared with the classical spin wave \cite{joliceour1990} values                 
with the spiral-collinear transition occurring at: $J_3=1/8$.
The direct spiral-collinear transition survives up to: $J_2 \gtrsim 1.8$, at which 
a disordered spin liquid is stabilized between the $(Q,Q)$-spiral and $(0,\pi)$-collinear 
order. 

\begin{figure*}
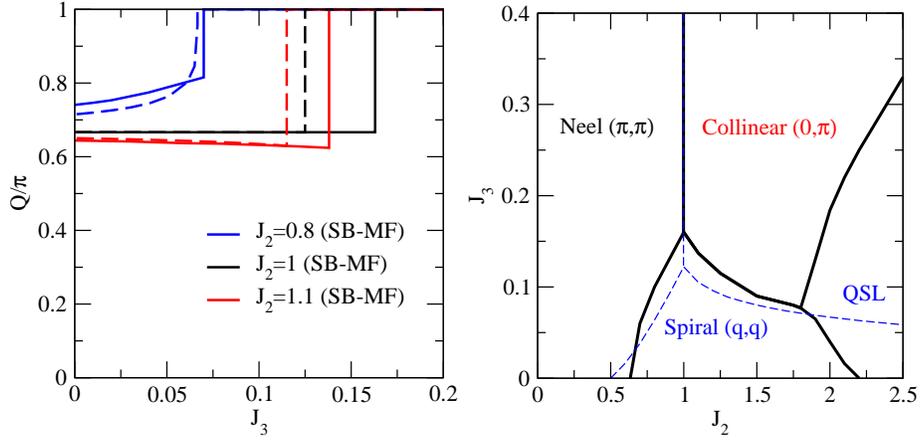

\epsfig{file=fig5a.eps,width=6cm,angle=0,clip=}
\epsfig{file=fig5b.eps,width=6cm,angle=0,clip=}
\caption{(Color online) Ordering wavevector and Schwinger boson mean-field phase diagram of 
the $J_1-J_2-J_3$ model on an anisotropic triangular lattice. In the left panel, the dependence of ${\bf Q}$ on $J_3$ 
showing the transition from the spiral: $(Q,Q)$, to the collinear
$(0,\pi)$ phase for which we set $Q=\pi$ for the purposes of this figure 
from SB-MF (solid lines) is compared to the classical wavevector dependence. The ground state phase 
diagram obtained from SB-MF (solid lines) is compared with the classical phase diagram (dashed lines) 
in the right panel showing the parameter range in which the spin liquid phase (QSL) is stable.}
\label{fig:phasedJ1J2J3J4}
\end{figure*}

The dependence of the ordering wave vector, ${\bf Q}$, with $J_3$ is shown 
in Fig. \ref{fig:phasedJ1J2J3J4} for different $J_2$. The absolute value of $Q$ in ${\bf Q}=(Q,Q)$ is plotted
as a function of $J_3$ until the jump to the $(0,\pi)$ phase occurs showing the 
discontinuous behavior of the order parameter signalling the first order transition.   For comparison, 
we plot the dependence of the classical ordering wavevector as a function of $J_3$ showing how
the transition point $(J_3)_c$ is shifted to larger values by the quantum fluctuation effects. 
Also it shows how the SB-MF ordering vector is enhanced with respect to the classical ordering vector 
for $J_2 < 1$ and is reduced when $J_2 >1$ independent of the value of $J_3$.
For $J_2=1$, the SB-MF ordering vector $Q=2 \pi/3$ is identical to the classical wave vector. 
Our results extend previous studies for the anisotropic triangular case with: $J_3=0$.

We summarize the results of ground state properties 
of the $J_1-J_2-J_3$ model of Fig. \ref{fig:phasedJ1J2J3J4} in which the SB-MF phase 
diagram is compared with the classical phase diagram in Fig. \ref{fig:phasedclass}.

\section{Magnetic excitations}
\label{sec:spinons}

We now discuss the elementary excitations of the system in different 
parameter regimes. Magnetically ordered states can be described with
the infinite lattice version of the SB-MF with the extra condition:
$\omega(\pm {\bf Q}/2)=0$. These type of solutions are recovered in large but
finite lattices by using equations (\ref{eq:mf}) with no extra 
condition. These solutions do not break the spin symmetry 
but have dispersions with a minimum energy which behaves as: 
$\omega(\pm {\bf Q}/2) \propto 1/N_s$. Disordered phases 
preserving the SU(2) spin symmetry of the hamiltonian are described through 
the version of the SB-MF approach expressed in Eq. \ref{eq:mf}.

\subsection{Elementary excitations: one-spinon dispersions}

The elementary excitations in the spin liquid phase described through SB-MF 
are the $S=1/2$ spinons. These can be visualized within Anderson "resonant valence bond"
(RVB) theory as $S=1/2$ defects propagating in the background of
resonating singlets covering the rest of the lattice. The SB-MF theory presented here
including singlet $A_{ij}$ and triplet $B_{ij}$ correlations corresponds to a 
large-$N$ saddle point\cite{flint2009} which appropriately deals with the
time-reversal properties of the spin in contrast to previous Sp$(N)$ theories\cite{sachdev1992}. 
At the large-$N$ saddle point or the SB-MF theory (for $N=2$) presented here, 
spinons are non-interacting. 

\begin{figure*}
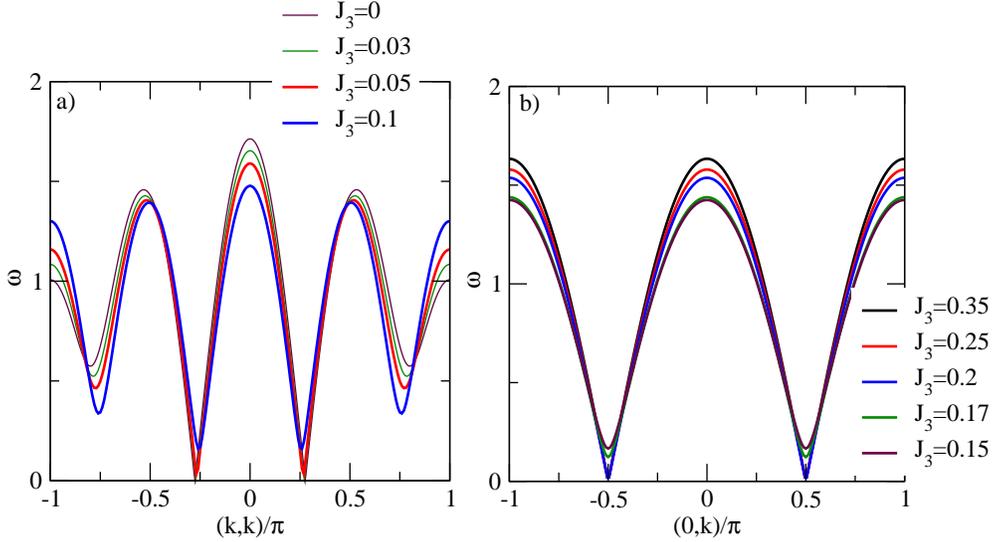

\epsfig{file=fig6a.eps,width=6cm,angle=0,clip=}
\epsfig{file=fig6b.eps,width=7.1cm,angle=0,clip=}
\caption{(Color online) Spinon dispersion for the $J_1-J_2-J_3$ model 
with $J_2=2$ and various $J_3$. In a) we show the evolution of the spinon dispersion along the 
diagonal $(k,k)$-direction of the Brillouin zone from the magnetically ordered 
spiral-$(Q,Q)$ phase to the spin liquid phase. In b) the evolution of
the spinon dispersion in the $(0,k)$ direction from 
the collinear-$(0,\pi)$ to the spin liquid is shown. The spin liquid phases are 
characterised by the opening of a gap.}
\label{fig:onespinonspiral}
\end{figure*}

The evolution of the one-spinon dispersion starting from the ordered $(Q,Q)$ spiral
phase as $J_3$ is increased with $J_2=2$ is shown in Fig. \ref{fig:onespinonspiral}.
Initially when $J_2=0$, the spinon dispersion is gapless and the 
spinons are Bose condensed at the $\pm {\bf Q}/2$ wave vectors leading to a 
small but finite magnetization. As $J_3$ is increased
and the disordered spin liquid reached the spinon dispersion develops a gap at 
$\pm {\bf Q}/2$ and long range order is lost.

The evolution of the spinon dispersions starting from the collinear-$(0,\pi)$
phase is also shown in Fig. \ref{fig:onespinonspiral} showing how the 
the gap opens at $(0, \pm \pi/2)$ on entering the spin liquid phase.

\subsection{Two-spinon excitations}

The $S=1/2$ Heisenberg antiferromagnetic chain with nearest-neighbor interaction, $J$, 
has no long range magnetic order and no energy gap
to the lowest excitation dispersion\cite{cloizeaux1962}: $\omega_q = {\pi J \over  2} |\sin q|$. 
Hence, magnetic excitations consist of a two-spinon continuum different from the well defined
dispersion of magnons, the 
magnetic quasiparticles expected in a three-dimensional ordered Heisenberg antiferromagnet. 

Within SB-MF, the triplet excitations can be formed by the composition of 
two spin-$1/2$ deconfined spinons. These form a broad particle-hole continuum
which reaches high energies with the minimum excitation energy related to 
magnetic order in ordered phases. In an ordered 
phase the lowest magnetic excitations are obtained by creating a spinon in 
the condensate and another spinon in the continuum.
The minimum two-spinon excitation energies read: 
\begin{equation}
\epsilon^{\pm}_{\bf k}=\omega_{\mp {\bf Q}/2}+\omega_{{\bf k} \pm {\bf Q}/2},
\label{eq:twospinondispersion}
\end{equation}
where $\omega_{\pm {\bf Q}/2} \rightarrow 0$, and $\epsilon^{\pm}_{\bf k}=\omega_{{\bf k} \pm {\bf Q}/2}$ 
in an ordered phase. 

\begin{figure*}
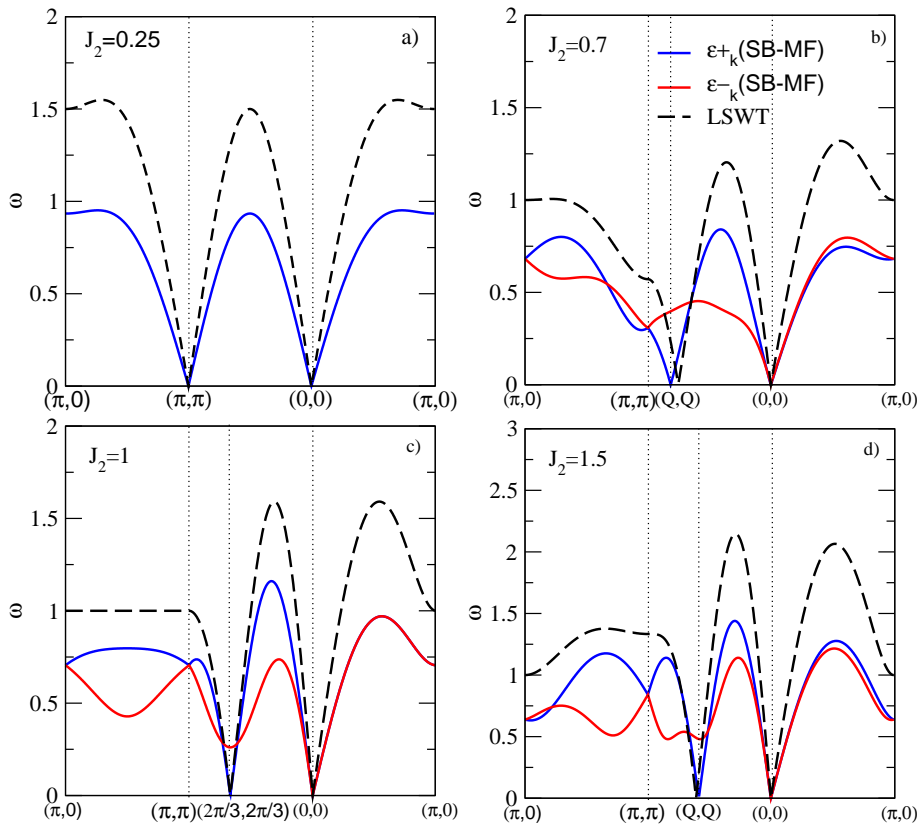

\epsfig{file=fig7a.eps,width=6cm,angle=0,clip=}
\epsfig{file=fig7b.eps,width=6cm,angle=0,clip=}
\epsfig{file=fig7c.eps,width=6cm,angle=0,clip=}
\epsfig{file=fig7d.eps,width=6cm,angle=0,clip=}
\caption{(Color online) Evolution of lowest two-spinon continuum energies from Schwinger boson mean-field theory
on an anisotropic triangular lattice ($J_3=0$). The blue and red full lines correspond to the
$\epsilon_{\bf k\pm{\bf Q}/2}$, in Eq. \ref{eq:twospinondispersion}, respectively. We show results within the 
a) N\'eel phase with $J_2=0.25$, b) spiral state with $J_2=0.7$, c) isotropic 
triangular lattice and d) $J_2=1.5$.  All these cases correspond to magnetically ordered phases.
The LSWT magnon dispersions are shown for comparison (dashed lines). Note that 
in the N\'eel phase the $\epsilon_{\bf k\pm{\bf Q}/2}$ excitations coincide. }
\label{fig:dispersion}
\end{figure*}

In Fig. \ref{fig:dispersion} we show the minimum two-spinon 
excitation energies of the continuum, $\epsilon^{\pm}_{\bf k}$, as obtained from Eq. (\ref{eq:twospinondispersion}) on 
an anisotropic triangular lattice ($J_3=0$). This is plotted in Fig. \ref{fig:dispersion} for 
different $J_2$ and compared to magnons obtained from spin-wave theory. We show the evolution
of these dispersions when going from the N\'eel to the spiral phases including the 
isotropic triangular lattice case already discussed in the literature\cite{mezio2011}.

(i) {\it N\'eel phases:} In the N\'eel phases 
we find that the lowest SB-MF dispersions
are very similar to the conventional magnon excitations. This is shown in Fig. 
\ref{fig:dispersion}a) for $J_2=0.25$. The only 
important effect is the smaller width of the SB-MF dispersions as compared to the semiclassical
LSWT dispersion. This is due to renormalization effects since the SB-MF theory contains
static interaction effects\cite{auerbachbook}  in a similar way as 
Hartree-Fock theory contains band renormalization and band shift effects in interacting electron 
models. Series expansion calculations in this
regime have found the development of 'roton' minima\cite{zheng2006} around $(\pi,0)$ 
in the $(\pi,0) \rightarrow (\pi/2,\pi/2)$ direction with a lower energy at $(\pi,0)$  
with respect to the $(\pi/2,\pi/2)$ wavevector. Both LSWT and SB-MF disagree 
with the series expansion result which predict a flat dispersion between these 
wavevectors.  A simple interpretation in terms of non-interacting spinons for these 
'roton' minima does not seem adequate and one should possibly need to
go beyond the mean-field theory and include spinon-spinon interactions.
  
(ii) {\it Spiral phases:} When entering the spiral phase we find the 
strongest deviations of the dispersions with respect to the LSWT. 
Already for $J_2=0.7$ (Fig. \ref{fig:dispersion}b) we find that 
apart from the renormalization effects discussed above there are also qualitative differences in the
momentum dependence in the $(\pi,0)$-$(\pi,\pi)$ direction. In the isotropic triangular case  
there is a flat band dispersion between $(\pi,0) \rightarrow (\pi,\pi)$ in LSWT which 
is not observed in the SB-MF dispersion but rather a minimum (maximum) occurs 
in the lowest (highest) branch at $(\pi,\pi/2)$ and dips at the 
$(\pi,\pi)$ and $(\pi,0)$ point which compare well with the roton minima observed in the 
series expansion results \cite{mezio2012}. This minima
can be associated with the existence of $(\pi,\pi)$-N\'eel and
$(0,\pi)$-collinear correlations\cite{mezio2011} in the $(Q,Q)$-spiral ordered phase. 
For larger $J_2$ the differences with the spin wave 
dispersion become more pronounced particularly around $(\pi,\pi)$  
where a deeper dip is observed compared to the 
LSWT magnon dispersions as in Fig. \ref{fig:dispersion}d).  

(iii) {\it Spin liquid formation:} We now discuss the evolution in 
the large $J_2$ limit where a spin liquid phase occurs. 
In Fig. \ref{fig:twospinon} we fix $J_2=2$ and 
increase $J_3$ so that we eventually enter the 
spin liquid phase. For $J_3=0.05$ the system is already in the spin liquid 
phase and a small gap opens in the dispersion around the short range ordering
spiral vector $(Q,Q)$. Concomitanly there is also a change in the 
momentum dependence of the dispersion with supression of the
dispersion at $(\pi,0)$ as compared to the $J_3=0$ case which indicates
 the proximity to the collinear phase. This is consistent with the expected 
behavior as extracted from the phase diagram (see Fig. \ref{fig:phasedJ1J2J3J4}).  
We also show in Fig. \ref{fig:twospinon}, the spin liquid obtained for $J_3=0.1$.


The above low energy magnetic dispersions will be modified in general 
in the presence of finite-$N$ fluctuations around the saddle point.
These generate gauge interactions that bind the spinons which in the ordered phases lead
to magnons in the neighborhood of the Goldstone modes. On the other hand,
at high-energy, pairs of spinons remain weakly bound.             

\subsection{Dynamical magnetic correlations}

The dynamics of the spin correlations in the system can be analyzed through 
inelastic magnetic neutron scattering experiments
which probe the $\Delta S= \pm 1$ excitations. If there are magnons
present in the magnetic excitation spectra, as in conventional magnets, then sharp quasiparticle peaks 
are found in the neutron scattering spectra. Since spinons carry half of the local 
spin degree of freedom at each lattice site, $\Delta S= \pm 1$, magnetic excitations
observed in neutron scattering can occur from the triplet combination of two spinons. 
Within SB-MF, spinons are deconfined leading to a two-spinon continuum  
rather than the sharp magnon quasiparticle peaks of conventional magnets. The dynamical spin correlation function
obtained in the SB-MF then reads:     
\begin{equation}
S^{zz}({\bf k},\omega)=\sum_n |\langle O|S^z_{\bf k}|n \rangle |^2 \delta(\omega -(E_n-E_{0})),
\end{equation}
with $S^z_{\bf k}=\sum_i e^{i {\bf k} \cdot {\bf R}_i} S^z_i $, and 
$S^z_i={1 \over 2} \left( a^\dagger_{i\uparrow} a_{i\uparrow}-a^\dagger_{i\downarrow} a_{i\downarrow} \right)$.
We evaluate this expression at the mean-field level using the Schwinger
boson approach. The ground state is defined as the vacuum of Bogoliubov
quasiparticles: $\alpha_{{\bf k}\sigma}|0 \rangle = 0$ where $\alpha_{{\bf k}\sigma}$
creates a Bogoliubov quasiparticle for any ${\bf k}$
and $\sigma$ as in Eq. (11). Excitation $n$ is produced by creating two spinons
above the vaccuum.  

Expressing the original boson operators in terms of the Bogoliubov quasiparticles
with the two-spinon excitations:
$E_n-E_{GS}=\omega_{\bf k_1}+ \omega_{-({\bf q}+{\bf k}_1)}$, the final expression for the
spin correlation function reads:
\begin{eqnarray}
S^{zz}({\bf k}, \omega)&=&{1 \over 4 N_s} \sum_{{\bf k_1}} | u_{{\bf k+k_1}} v_{{\bf k_1}}
-u_{\bf k_1} v_{\bf k+ k_1}|^2 \notag\\&&\times \delta(\omega -(\omega_{-\bf k_1}+\omega_{\bf k+k_1})),
\label{eq:szz}
\end{eqnarray}
with the matrix elements: $u_{\bf k}=\sqrt{(1+{B({\bf k})+\lambda \over \omega_{\bf k} })/2}$
and $v_{\bf k}=i \textrm{sign}(A({\bf k}))\sqrt{(-1+{B({\bf k})+\lambda \over \omega_{\bf k} })/2}$.
The above Eq. (\ref{eq:szz}) gives the spectra of $S=1$ excitations relevant to
neutron scattering consisting on two spinons.
The lowest energy particle-hole processes described by $S^{zz}({\bf k}, \omega)$ 
 correspond to exciting a spinon in the condensate and another one in the continuum.                      
For finite size lattices: $(B({\bf Q})+\lambda)/\omega({\bf Q}/2)=N_s m({\bf Q})$,
$u_{\bf \pm Q/2} \sim \sqrt{N_s  m({\bf Q})/2}$ and 
$v_{\bf \pm Q/2} \sim i  \sqrt{N_s  m({\bf Q})/2} $ and so the weight right at $\pm {\bf Q/2}$  
is proportional to the magnetization. 

\begin{figure*}
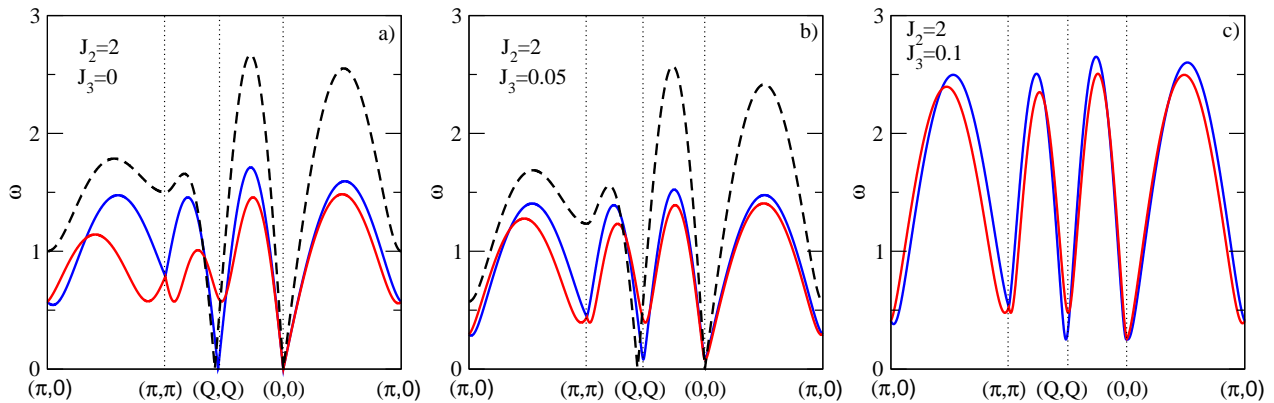

\epsfig{file=fig8a.eps,width=5.5cm,angle=0,clip=}
\epsfig{file=fig8b.eps,width=5.5cm,angle=0,clip=}
\epsfig{file=fig8c.eps,width=5.5cm,angle=0,clip=}
\caption{(Color online) Evolution of the lowest energies of the two-spinon continuum from Schwinger boson mean-field theory
on the $J_1-J_2-J_3$ for $J_2=2$ and different $J_3$. The plots and labels are the same as in Fig. \ref{fig:dispersion}.
In a) for $J_3=0$, the system is in an ordered spiral phase  whereas in b) for $J_2=0.05$, the
system is a spin liquid characterised by a small energy gap, $\Delta \approx 0.08$ and short range spiral
$(Q,Q)$ correlations. In c) the same plots for $J_3=0.1$ showing the enhancement of the gap.
}
\label{fig:twospinon}
\end{figure*}

\begin{figure*}
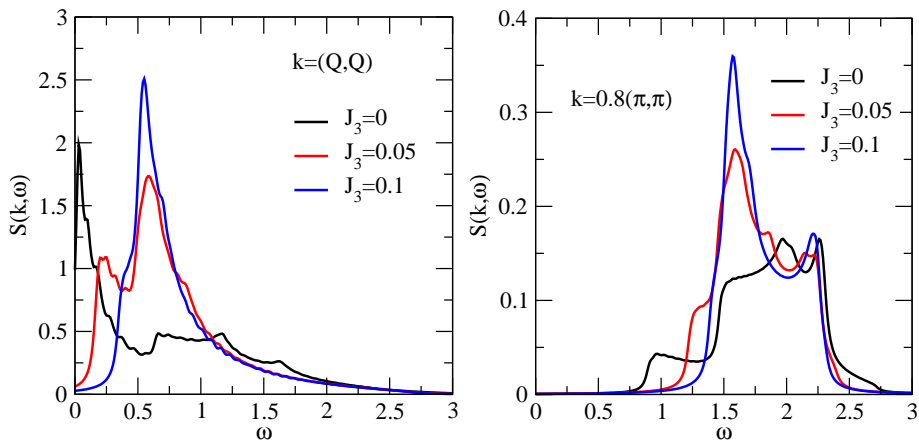

\epsfig{file=fig9a.eps,width=6cm,angle=0,clip=}
\epsfig{file=fig9b.eps,width=6cm,angle=0,clip=}
\caption{(Color online) Dynamical spin structure factor, $S^{zz}({\bf k}, \omega)$,
for $J_2=2$. In the left panel we show $S^{zz}({\bf k}, \omega)$
at the spiral ordering wave vector ${\bf Q}$ whereas in the right panel we 
use ${\bf k}=0.8(\pi,\pi)$. The plot shows
how the two-spinon continuum described by Schwinger boson mean-field theory 
reaches high energies of about $2.5-3$ which is an artifact of the approach.\cite{mezio2011}
Note the much smaller vertical scale in the right
plot ${\it i. e.}$ away from the ordering wave vector, ${\bf Q}$.
}
\label{fig:twospinonspectra}
\end{figure*}

In Fig. \ref{fig:twospinonspectra} the dynamical spin correlations, 
$S^{zz}({\bf k}, \omega)$ are shown for $J_2=2$ going from the 
spiral ordered phase into the spin liquid 
phase corresponding to parameters shown in Fig. \ref{fig:twospinon}.  
The main features observed in the spectra correspond to the elementary
two-spinon branches: $\epsilon^{\pm}_{\bf k}$ plotted in Fig. \ref{fig:twospinon}. 
We first concentrate in $S({\bf k}, \omega)$, evaluated at the 
ordering wave vector: ${\bf k}=(Q,Q)$.
In the ordered phase for $J_3=0$, there is a very low energy peak (going to zero 
in the infinite system) which dominates the 
spectra and corresponds to the Goldstone mode associated
with the long range spiral magnetic order. A second smaller feature occurs at the
second elementary branch of Fig. \ref{fig:twospinon}.
Apart from these two main features, there is a contribution of particle-hole excitations
which extends up to high energies. Such contribution is associated with 
two-spinon excitation processes involving spinons in the normal 
fluid (not condensed) as recently pointed out\cite{mezio2011}. As $J_3$ is increased, there is a redistribution of spectral
weight. On entering the spin liquid phase, a gap opens up in the spectra and the 
spectral weight of the lowest branch is suppressed while there is an enhancement
of spectral weight of the highest magnetic excitation.  
For the wave vector ${\bf k}=0.8(\pi,\pi)$ different from ${\bf Q}$, there is 
also a two-peak structure similar to the one discussed above associated with $\epsilon^{\pm}_{\bf k}$. 
However, the overall spectral weight contribution is suppressed as compared to ${\bf k}=(Q,Q)$ 
since excitations have higher energy. 

\section{Conclusions}
\label{sec:conclusions}

We have analyzed the effect of a third-nearest neighbor antiferromagnetic interaction, $J_3$,
on the magnetic properties of the antiferromagnetic Heisenberg model on an 
anisotropic triangular lattice. We have shown that $J_3$ can frustrate the long range 
spiral magnetic order leading to a spin liquid phase when 
$J_2>1.8$ and a small $J_3 \lesssim 0.1$.  Since SB-MF is known to favor ordered states \cite{auerbachbook} the 
parameter regime in which the spin liquid phase is stable may be enlarged 
by fluctuations. 

The antiferromagnetic coupling, $J_3$, considered here may be generated through either, second order, superexchange
processes between third nearest-neighbors sites or fourth order processes, that also drive ring exchange. Ring exchange involving
four sites can be separated into two-spin $J_3$ Heisenberg and four-spin \cite{liming2000,michael2013}
contributions. Our present analysis, focusing on the frustrating effects associated with the
former Heisenberg-type exchange terms in the ring exchange, is helpful in the 
understanding of ring exchange effects in frustrated antiferromagnets.

Our analysis may be relevant to recent observations suggesting spin liquid behavior in 
certain layered materials.
 Cs$_2$CuBr$_4$  is an anisotropic triangular material with $J_2 \sim 2$ 
which would be predicted to be magnetically ordered. However, from our analysis
a rather small $J_3$ would be sufficient to turn it into a spin liquid.        
On the other hand, Cs$_2$CuCl$_4$ with $J_2 \sim 3$ would be 
a spin liquid from our analysis even for $J_3=0$ which is in contrast with series
expansion predictions. In any case, our SB-MF analysis suggests that in these
two materials $J_3$ may play a role in determining their magnetic properties
\cite{coldea2001}. On the other hand, organic materials in which spin liquid phases have been 
found such as $\kappa$-(BEDT-TTF)$_2$Cu$_2$(CN)$_3$ and Me$_3$EtSb[Pd(dmit)$_2$]$_2$, are in a different parameter regime  
$J_2 \sim 0.7$ in which SB-MF would predict an ordered state regardless of $J_3$ unlike the 
spin liquid predicted by series expansions \cite{zheng1999} for $J_3=0$. On the basis of 
this work, one would expect a finite $J_3$ to further stabilize the QSL. It would be 
interesting to find organic materials which are in the large $J_2$ parameter regime 
discussed here as they would be strong candidates for the observation of spin liquid behavior.      

The SB-MF prediction for the magnetically disordered state 
 is a $Z_2$ spin liquid\cite{sachdev1992,zhou2003} characterized by gapped bosonic excitations.
However, the $T$-dependence of the NMR relaxation rate in Cs$_2$CuCl$_4$
and $\kappa$-(BEDT-TTF)$_2$Cu$_2$(CN)$_3$ suggests the presence of gapless excitations
in the system. This fact can be more naturally explained in terms of
fermionic mean-field theories with a ground state consisting of a spinon Fermi surface\cite{motrunich2005}
but is not inconsistent with a bosonic mean-field state of the type described here with 
spin singlet-triplet gaps which are smaller\cite{marston2012} than $J_1/10$. Further theoretical efforts
should concentrate in understanding these observations by going beyond the mean-field
theory used here using numerical techniques that can treat the constraint on the number of bosons exactly.

\appendix

\section{Classical energy}
Here, we analyze how in the $S \rightarrow \infty $ limit the ground state
energy obtained from SB-MF converges to the classical ground state energy \cite{gazza1993}. 
Self-consistent solutions of the bond strengths of the model in the classical limit are given by:  
\begin{eqnarray}
B_{ij} \approx S\cos({\bf Q}\cdot {\bf R}_{ij}/2) \\ \nonumber
A_{ij} \approx S\sin({\bf Q}\cdot {\bf R}_{ij}/2) \\ \nonumber 
\end{eqnarray}
where ${\bf R}_{ij}$ is the distance between two sites forming a bond. 
One can check that the classical energy for a given bond is indeed recovered:
\begin{equation}
\langle {\bf S}_i \cdot {\bf S}_j \rangle = |B_{ij}|^2 -|A_{ij}|^2 \approx S^2 \cos({\bf Q} \cdot {\bf R}_{ij}).
\end{equation}

The boson chemical potential in the magnetically ordered phase is then given by:
\begin{equation}
\lambda=A({\bf Q}/2)-B({\bf Q}/2)=-S J({\bf Q })-S^3 J_{ring}({\bf Q})=-{E_{class} \over S},  
\end{equation}
where we have defined:
\begin{eqnarray} 
&&J({\bf Q })=J_1(\cos(Q_x)+\cos(Q_y))+J_2\cos(Q_x+Q_y) \\  
\nonumber
&&+J_3 (\cos(Q_x-Q_y)+\cos(2 Q_x+Q_y)+ \cos(Q_x+2 Q_y)), 
\end{eqnarray}
and the classical energy: $E_{class}=S^2 J({\bf Q})$.
The mean-field energy referred to the chemical potential obtained from $H^{MF}$ reads:
\begin{equation}
E= \langle H^{MF} \rangle +2\lambda S={1 \over N_s}\sum_{\bf k} \omega_{\bf k}
- S^2J({\bf Q})\approx E_{class}, 
\end{equation}
when $S$ is large since the sum over $\omega_{\bf k}$ in the right hand 
side of the equation is of O($S$) only. Therefore, when $S \rightarrow \infty $ 
the SB-MF energy converges to the classical energy.

\begin{acknowledgments}
This work was funded in part by the Australian Research Council under the Discovery (DP1093224),
Future (FT130100161) and
QEII (DP0878523) schemes. J. M. acknowledges financial support from MINECO (MAT2012-37263-C02-01).
\end{acknowledgments}


\begin{thebibliography}{39}

\bibitem{balents2010} L. Balents, Nature, {\bf 464},  199 (2010).
\bibitem{ashcroftbook} N. Ashcroft and D. Mermin, {\it Solid State Physis}, (Thomson-Learning, 1975).
\bibitem{Tsvelik} A. Tsvelik, {\it Quantum Field Theory in Condensed Matter Physics } (Cambridge University Press, 2003).
\bibitem{Janani} C. Janani, J. Merino, I. P. McCulloch, B. J. Powell, arXiv:1401.6605.
\bibitem{scriven2012} E. P. Scriven and B. J. Powell, Phys. Rev. Lett. {\bf 109}, 097206 (2012).
\bibitem{john2007} J. O. Fjaerestad, {\it et. al.}, Phys. Rev. B {\bf 75}, 174447 (2007).
\bibitem{shirata2012}  Y. Shirata, H. Tanaka, A. Matsuo, and K. Kindo, Phys. Rev. Lett. {\bf 108}, 057205 (2012).
\bibitem{zhou2011} H. D. Zhou, {\it et. al.} , Phys. Rev. Lett. {\bf  106} , 147204 (2011).
\bibitem{susuki2013} T. Susuki, {\it et. al.}, Phys. Rev. Lett. {\bf 110}, 267201 (2013).
\bibitem{RPP} B. J. Powell and R. H. McKenzie
Rep. Prog. Phys. {\bf74}, 056501 (2011).
\bibitem{sachdev1992} S. Sachdev, Phys. Rev. B {\bf 45},  12377 (1992).
\bibitem{motrunich2005} O. I. Motrunich, Phys, Rev. B {\bf 72}, 045105 (2005).
\bibitem{shimizu2006} Y. Shimizu, K. Miyagawa, K. Kanoda, M. Maesato, and G. Saito, Phys, Rev. B {\bf 73} 140407 (2006).
\bibitem{Normand} B. Normand, Contemporary Phsyics {\bf50}, 4 533 (2009).
\bibitem{marston2012}  M.-A. Vachon {\it et. al.},  New J. Phys. {\bf 13}, 093029 (2011).
\bibitem{balents2007} M. Kohno, O. A. Starykh and L. Balents, Nat. Phys. {\bf 3}, 790 (2007).
\bibitem{coldea2001} R. Coldea, D. A. Tennant, A. M. Tsvelik, and Z. Tylczynski, Phys. Rev. Lett. {\bf 86} 1335 (2001);
R. Coldea, D. A. Tennant, and Z. Tylczynski, Phys. Rev. B {\bf 68} 134424 (2003).
\bibitem{elstner1993} N. Elstner, R. R. P. Singh, and A. P. Young, Phys. Rev. Lett. {\bf 71} 1629 (1993).
\bibitem{bernu1994} B. Bernu, P. Lecheminant, C. Lhuillier, and L. Pierre, Phys. Rev. B {\bf 50} 10048 (1994).
\bibitem{anderson1973} P. W. Anderson, Mater. Res. Bull. {\bf 8}, 153 (1973).
\bibitem{wang2006}F. Wang and A. Vishwanath, Phys. Rev. B 74, 174423 (2006).
\bibitem{liming2000} W. LiMing, G. Misguich, P. Sindzingre, and C. Lhuillier, Phys. Rev, B {\bf 62}, 6372 (2000);
G. Misguich, C. Lhuillier, B. Bernu, and C. Waldtmann, Phys. Rev. B {\bf 60}, 1064 (1999).
\bibitem{michael2013} M. Holt, B. J. Powell, and J. Merino, arXiv:1310.4597.
\bibitem{cepas2010} L. Messio, O. C\'epas, and C. Lhuillier, Phys. Rev. B {\bf 81}, 064428 (2010).
\bibitem{auerbachbook} A. Auerbach, {\it  Interacting electrons and quantum magnetism}, Springer-Verlag (1994).
\bibitem{gazza1993}  C. J. Gazza and H. A. Ceccatto, J. Phys: Condens. Matter {\bf 5}  L135 (1993).
\bibitem{mezio2012} A. Mezio, L. O. Manuel, R. R. P. Singh and A. E. Trumper, New Journal of Physics. {\bf 14} 123033 (2012).
\bibitem{flint2009} R. Flint and P. Coleman, Phys. Rev. B {\bf 79}, 014424 (2009).
\bibitem{bursill2006} M. Q. Weng, D. N. Sheng, Z. Y. Weng, and R. J. Bursill, Phys. Rev. B {\bf 74}, 012407 (2006).
\bibitem{trumper1999}  A. E. Trumper, Phys. Rev. B, {\bf 60} 2987 (1999).
\bibitem{merino1999} J, Merino, R. H. McKenzie, J. B. Marston, and C. H. Chung, J. Phys. Condens. Matter {\bf 11}, 2965 (1999).
\bibitem{hauke2013} P. Hauke, Phys. Rev. B {\bf 87}, 014415 (2013).
\bibitem{zheng1999} Zheng Weihong, Ross H. McKenzie, and R. R. Singh, Phys. Rev.  B {\bf 59}, 14367 (1999).
\bibitem{manuel1999} L. O. Manuel and H. A. Ceccatto, Phys. Rev. B, {\bf 60} 489 (1999).
\bibitem{chunghou2001} C. H. Chung, J. B. Marston, and Ross H. McKenzie, J. Phys.: Condens. Matter {\bf 13}, 5159   (2001).
\bibitem{tocchio2013} L. F. Tocchio. H. Feldner, F. Becca, R. Valenti, and C. Gross, Phys. Rev. B {\bf 87}, 035143 (2013).
\bibitem{joliceour1990} Th. Joliceour, E. Dagotto, E. Gagliano, and S. Bacci, Phys. Rev. B {\bf 42}, 4800 (1990).
\bibitem{read1991} N. Read and S. Sachdev, Phys. Rev. Lett. {\bf 66}, 1773 (1991).
\bibitem{ceccatto1993} H. A. Ceccatto, C. J. Gazza, and A. E. Trumper, Phys. Rev. B {\bf 47}, 12329 (1993).
\bibitem{lefmann1994} K. Lefmann and P. Hedegard, Phys. Rev. B {\bf 50}, 1074 (1994).
\bibitem{mezio2011} A. Mezio, {\it et. al.}, Europhysics Letters, {\bf 94} 47001 (2011).
\bibitem{cloizeaux1962} J. des Cloizeaux and J. J. Pearson, Phys. Rev. B {\bf 128}, 2131 (1962).
\bibitem{zheng2006} Zheng Weihong, {\it et. al.},  Phys. Rev. B {\bf 74}  224420 (2006). 
\bibitem{zhou2003} Y. Zou and X.-G Wen, cond-mat/0210662.
\end{thebibliography}
\end{document}